\documentclass[%
 reprint,
 amsmath,amssymb,
 aps,
]{revtex4-1}

\usepackage{graphicx}
\usepackage{dcolumn}
\usepackage{bm}
\usepackage{hyperref}
\usepackage{wasysym}

\begin{document}

\preprint{APS/123-QED}

\title{Interactions between  colloids induced by a soft cross-linked polymer substrate}

\author{Lorenzo Di Michele}
 \author{Taiki Yanagishima}
 \author{Anthony R. Brewer}
 \author{Jurij Kotar}
 \author{Erika Eiser}
 \email{ee247@cam.ac.uk}
 \affiliation{Biological and Soft Systems, Cavendish Laboratory, University of Cambridge,\\J.J. Thomson Avenue, Cambridge, CB3 0HE United Kingdom}
 \author{Seth Fraden}
 \affiliation{Brandeis University, Martin Fisher School of  Physics, Waltham, MA 02454 USA}




\date{\today}
\begin{abstract}
Using video-microscopy imaging we demonstrate the existence of a  short-ranged equilibrium attraction between heavy silica colloids diffusing on  soft surfaces of cross-linked polymer gels. The inter-colloid potential can be tuned by changing the gel stiffness or by coating the colloids with a polymer layer. On sufficiently soft substrates, the interaction induced by the polymer matrix leads to large-scale colloidal aggregation.  We correlate the in-plane  interaction with a colloid-surface attraction.
\begin{description}
\item[PACS numbers] {82.70.Dd, 82.35.Lr, 68.08.-p}
\end{description}
\end{abstract}

\maketitle
Knowledge of the interactions between colloids confined to surfaces~\cite{Seth2,Grier2007,Squires2000} or interfaces~\cite{Dietrich2005} is key to the fundamental understanding of many physical phenomena. For instance, interaction between colloids on surfaces can induce the self-assembly of ordered phases~\cite{Im2002} that may find applications in the engineering of photonic crystals~\cite{Xia2000,Im2002,Tarhan1996,Xia2000}.  Several aspects of cellular morphology~\cite{Guck2010,Discher2005}, mechanical properties~\cite{Engler2004,Discher2005}, mobility~\cite{Ulrich2009,Discher2005} and cell differentiation~\cite{Zajac2008,Discher2005} have been found to be sensitive  to the elastic properties of the environment. Experiments in this field often employ cross-linked polymer-gel surfaces~\cite{Engler2004,Ulrich2009,Guck2010} as a model for biological tissues. We demonstrate that soft polymer substrates may actively induce non-negligible interactions between inanimate, micrometer-sized, objects. Analogous phenomena may affect as well the behavior of living systems of the same dimension and must be taken into account when the results of above cited experiments are analyzed. In this Letter we explore the nature of such substrate-induced inter-colloidal forces and show how they are modified by changing physical properties of the soft polymer gels.\\
The experiments were performed using plain silica colloids with diameter $\sigma= 1.16\pm0.05$ $\mu$m (Microparticles GmbH, Berlin) and a nominal density of 1.5-2 g cm$^{-3}$ suspended in 100 mM tris(hydroxymethyl)aminomethane (TRIS) buffer at pH 8. Under these conditions the silica colloids have a net negative charge with a $\zeta$-potential of -42$\pm1$ mV. As soft surfaces we used 150 $\mu$m thick polyacrylamide (PAA) cross-linked gel films deposited on top of a microscope coverslip. 
The polymerization of the PAA was triggered by adding tetramethylethylenediamine and ammonium persulfate to a solution containing acrylamide monomers and the cross-linker bis-acrylamide (bis-AA) in phosphate buffered saline buffer (PBS, pH 7.4). 
For control experiments we coated the same silica colloids with positively charged poly-{\footnotesize{L}}-lisine-poly(ethylene glycol) (PLL-PEG, Surface Soultions, Switzerland) that is easily adsorbed onto silica and provides steric stabilization. PLL-PEG coated beads were almost neutral with a $\zeta$-potential of -2.2$\pm$0.2~mV. Non-adhesive hard surfaces were obtained by coating the glass bottom of 8 mm diameter incubation wells (Sensoplate, Greiner bio-one) with PLL-PEG. \\
\begin{figure}
\includegraphics{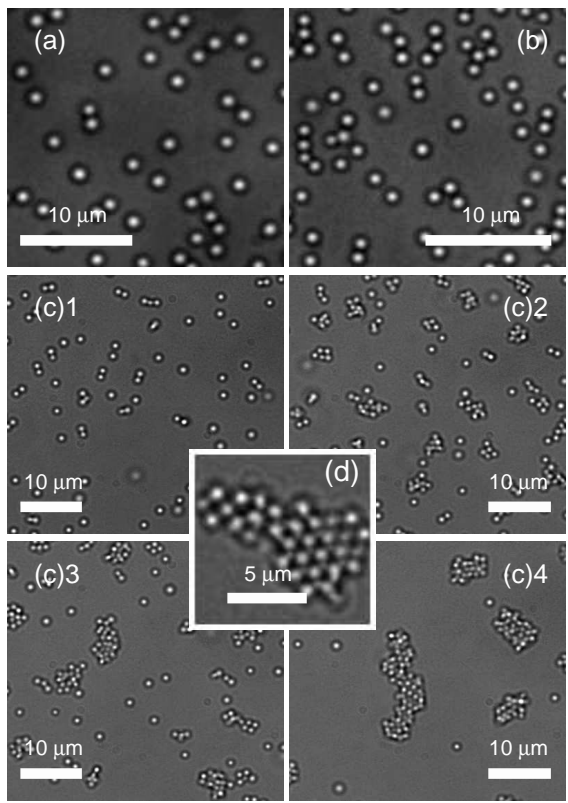}
\caption{\label{figure_beads} Panels (a) and (b): microscope images of plain silica colloids sedimented on a PAA soft gel with different surface coverage. Panels (c)1-4: aggregation process of plain silica colloids sedimented on ultra-soft PAA gel surface at $t=0$~(1), $t\approx15$ min~(2), $t\approx 30$ min~(3) and $t\approx60$ min~(4). Panel~(d): highlight of an ordered aggregate on the ultra-soft gel.}
\end{figure}
Using a PAA concentration of 5 \% we tuned the stiffness of the material by changing the bis-AA fraction. 
The observations of weak attractive  interactions between colloids were qualitatively and quantitatively similar  for gels with elastic shear modulus $G'$ \footnote{Measured using Physica MCR bulk rheometer, Anton Paar, Austria.} between 55 and 522 Pa, thus, we focus on a substrate  prepared with $4\times10^{-4}$ v/v bis-AA resulting in a $G'=240\pm5$ Pa --  we refer to this gel as the ``soft" substrate. In contrast,  colloids on ``ultra-soft'' gels obtained with bis-AA concentration of $1 \times 10^{-4}$ v/v and with $G'=12\pm3$ Pa exhibited stronger interactions.
On soft gels, sedimented colloids did not undergo large scale aggregation, however, we observed the formation of dimers, triplets, and small chains and clusters, as shown in Fig.~\ref{figure_beads}(a) and \ref{figure_beads}(b) for two different surface densities.
On the ultra-soft gel, sedimented beads initially formed small aggregates that were still mobile and  merged in larger ones. The aggregation stopped after about one hour as the reduced mobility of big aggregates kinetically hindered the process. A sequence of images of the aggregation is shown in Fig.~\ref{figure_beads}(c). The larger aggregates could be either amorphous or ordered in a hexagonal close-packed lattice [see  Fig.~\ref{figure_beads}(d)].  The mobility of colloids within the aggregates was small and we only observed rearrangement of the beads located at the outer perimeter of the clusters. Using PLL-PEG coated colloids instead of plain ones, the aggregation still took place on the same timescale but the clusters showed  liquid-like behavior with a continuous rearrangement of single colloids in the bulk of the aggregates.\\
\begin{figure}
\includegraphics{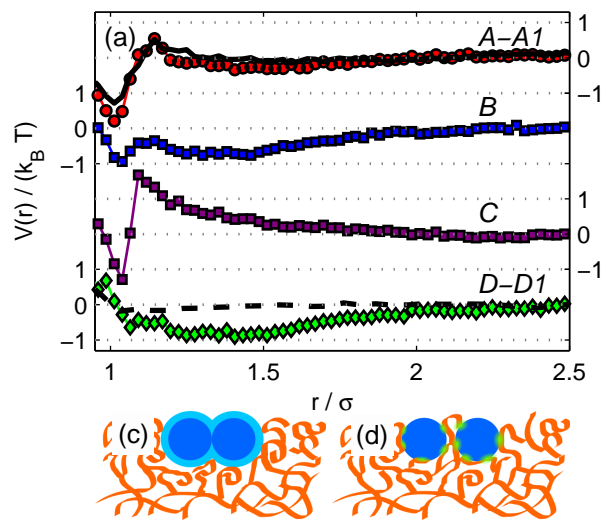}
\caption{\label{figure_pair} (color online) (a) Pair potentials of colloids sedimented on gels and glass surfaces measured with the blinking-tweezer (BT) or pair-distribution (PD) methods. Bare silica colloids sedimented on soft PAA gel surface measured with BT ({\it A} $\circ$) and with PD ({\it A1} solid line); PLL-PEG coated silica colloids sedimented on a soft PAA gel surface measured with BT ({\it B}); PLL-PEG coated silica colloids sedimented on a ultra-soft PAA gel surface measured with BT ({\it C}); PLL-PEG coated silica colloids sedimented on a PLL-PEG coated rigid glass surface measured with BT ({\it D}, $\diamond$) and with PD ({\it D1}, dashed line). The curves have been vertically shifted for clarity. Not-to-scale sketch of two beads partially included in the gel surface and interacting through excluded volume effects (b) and colloid-polymer adhesion (c).}
\end{figure}
We measured the effective pair-potential of sedimented beads following the method proposed by Crocker and Grier~\cite{Crocker1994} that is sometimes referred to as ``blinking optical tweezers" (BT)~\cite{Crocker1996b}. In the BT experiments, we positioned two isolated colloids at separations between 1.5 and 10 $\mu$m using optical tweezers \footnote{Home-made optical tweezers based on a 2W 1064nm laser (CrystaLaser) mounted on a Nikon Eclipse Ti-E microscope with a Plan Apo VC 60x WI 1.2 N.A.
objective and a Pike F-100B CCD camera (Allied Vision Technologies).}. Then, just after releasing the colloids, we took movies of 15 s at 30 frames per second. The separation $r$ between colloids was obtained by tracking their positions using conventional tracking algorithms \footnote{We developed our customized software based on open source MATLAB routines available at the URL \url{http://physics.georgetown.edu/matlab}}. After discretizing $r$ with a mesh size of 25 nm we sampled the transition matrix elements $P_{ij}$ that express the probability for the separation to evolve from the bin $i$ to the bin $j$ in the time between two consecutive frames. The stationary probability distribution of $r$ is $\rho^{\mathrm{s}}_i=P_{ij}\rho^{\mathrm{s}}_j$ and the effective pair-potential is given by  $V(r)/(k_{\mathrm{B}}T)=-\log\left(\rho^{\mathrm{s}}\right)$. The BT method has the virtue that it produces good statistics, but there are two known artifacts with this technique; optical interference~\cite{Baumgartl2005,Grier2007} and hydrodynamic interactions \cite{Squires2000}.  Superposition of colloid diffraction-images can result in errors in the determination of the inter-colloid distance $r$, and this can lead to systematic errors in the estimate of $V(r)$~\cite{Baumgartl2005,Grier2007}. We corrected our data for these errors following the method proposed by Polin et al.~\cite{Grier2007}.\\ 
For the case of plain silica colloids supported by a soft PAA gel surface, the apparent $V(r)$ shows two main features: a long-range attractive minimum with a depth of about $0.5~k_{\mathrm{B}}T$ and a sharp minimum for $r\approx\sigma$, as shown in Fig.~\ref{figure_pair}(a), curve {\it A}. Figure \ref{figure_pair}(a) curve {\it B} demonstrates that colloids covered with a PEG brush had a weakened short range attraction when compared to bare colloids on the same soft gel. In Fig.~\ref{figure_pair}(a) curve {\it C} we show $V(r)$ measured for PLL-PEG coated colloids on the ultra-soft gel \footnote{Tracking the distance between plain colloids on ultra-soft gels was not reliable due to rapid aggregation}. Here, the short-range attraction is stronger than for soft gels, while the long-range minimum is absent.
In Fig.~\ref{figure_pair}(a) curves {\it A}, {\it B} and {\it C} there is a maximum of the potential located at $r\,\approx\,1.2\,\sigma$. This cannot be ascribed to screened Coulomb repulsion as the Debye screening length is estimated to be only a few nanometers \footnote{For practical reasons our samples could not be sealed. The resulting slow evaporation could possibly increase the buffer concentration leading to further reduction in the Debye screening length. No measurable flow at the level of the surface was observed.}. As a control experiment we measured $V(r)$ with BT for the case of PLL-PEG coated colloids sedimented on a PLL-PEG coated rigid glass surface. The potential, plotted in Fig.~\ref{figure_pair}(a) curve {\it D}, does not show the short-range minimum observed on soft surfaces, although the long-range attraction is stronger.\\
Because the long-range attraction in the effective pair-potentials for colloids was observed for both soft gels and rigid glass surfaces, it must have the same origin. The form of the long-range attractive potential resembles the attraction observed between like-charged colloids in the presence of two confining walls~\cite{Crocker1996}. As pointed out by Squires and Brenner~\cite{Squires2000}, the potential extracted using BT method arises from forces which may not have an underlying equilibrium potential. In particular, a hydrodynamic coupling between two moving colloids in the presence of a confining wall may produce an apparent effective attractive interaction similar to the long-range attraction that we measured.\\ 
To separate this non-equilibrium hydrodynamic effect from the equilibrium potential between supported colloids, we also evaluated $V(r)$ from the radial distribution function $g(r)$ obtained by measuring positions of colloids in equilibrium conditions using artifact corrected video microscopy \cite{Grier2007}. For each sample we took a few hundred pictures at time intervals of 1 s of an area of about 1.5$\times10^{-2}$ mm$^2$ with a surface coverage of 0.5-1 \%. At these densities we can neglect many body effects and extract the pair-potential as $V(r)/(k_{\mathrm{B}}T)\approx -\log\left[g(r)\right]$. We determined $V(r)$ from measurements of $g(r)$ for the cases of plain colloids settled on a soft gel surface and PLL-PEG coated colloids settled on a rigid glass surface. The potentials are shown in Fig.~\ref{figure_pair}(a) curves {\it A1} and {\it D1} respectively. In both cases, the long-range minimum is absent, consistent with the explanation that this attraction was due to hydrodynamic coupling.
The variations that we observe in measured potential energy between ultra-soft gels, soft gels and rigid surface correlates with differences in the colloid-surface interaction, as well as to a penetration of the hydrodynamic flow in the polymer network.\\ The short-range attraction is present for the cases of the soft and ultra-soft gel surfaces, but is absent for the rigid glass surface as measured using both the non-equilibrium BT and equilibrium pair-distribution methods. These observations, taken together, indicate that the short-range colloid-colloid attraction is an equilibrium effect mediated by the polymer gel substrate. The same argument applies to the repulsive maximum in the colloid pair-potential located at $r\,\approx\,1.2\,\sigma$ that is present when the colloids are on soft gels, but absent when colloids are on a hard glass interface, as shown in Fig.~\ref{figure_pair}. We can deduce that the short-range attraction is responsible for the small aggregates on the soft gel substrates and for the large-scale clustering on the ultra-soft ones, where it became stronger. The weakening due to the PLL-PEG coating accounts for the liquid-like behavior of clusters on the ultra-soft gels.\\
The range and shape of the short-range attractive well are compatible with a description in terms of depletion~\cite{ASAKURA1954,*ASAKURA1958}. 
The PAA gel can in principle release free polymers in solution that may act as depletants. However, the absence of aggregation on bare glass in the same sample cell that contains gels with associating colloids excludes this possibility. Chen and Ma~\cite{Chen2005} predicted that a combination of excluded volume effects and polymer-colloids interfacial energy can give rise to an interaction between colloids embedded in a polymer brush. A similar effect can occur in our system assuming a partial inclusion of the colloids in a weakly cross-linked polymer layer on top of the substrates.\\
We explored this hypothesis by measuring the colloid-surface interaction energy.
\begin{figure}
\includegraphics{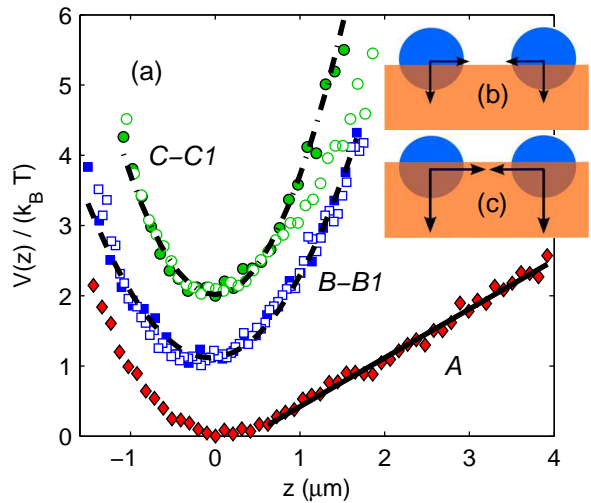}
\caption{\label{figure_zpot} (color online) (a) Experimental colloid-surface interaction potentials for PLL-PEG coated silica colloids confined by PLL-PEG coated glass surface ({\it A}), plain ({\it B}, filled symbols) and PLL-PEG coated ({\it B1}, empty symbols) silica colloids confined by a soft PAA gel surface; plain ({\it C } filled symbols) and PLL-PEG coated ({\it C1} empty symbols) silica colloids confined by an ultra-soft PAA gel surface. The solid line is a linear fit. Dashed and dot-dashed lines are quadratic fit. $z$ is the direction normal to the surface. The curves are shifted vertically for clarity. In the insets: sketch of coupling between surface-colloid and colloid-colloid interactions for soft (b) and ultra-soft (c) gels.}
\end{figure}
We tracked fluctuations of the distance $z$ of sedimented beads from the surface by video-microscopy measurements~\cite{Seth1}. The diffraction image of a colloid changes as a function of the distance between the bead and the focal plane of the objective lens. We took a set of a few hudred pictures of an immobile colloid by moving the focus with steps of 25 nm, then we associated an intensity $I(z)$ to each reference image by integrating the pixel value within a circle concentric with the bead. Fitting the intensity as a function of $z$ we obtained a calibration curve that is monotonic within an interval of a few microns. Working with the same illumination conditions we took movies of single freely diffusing colloids, then we measured $I(t)$ for each frame and found the corresponding $z(t)$ using the calibration curve. We sampled the probability distribution $\rho(z)$ by taking a histogram of $z(t)$. The colloid-surface potential is given by $V(z)/(k_{\mathrm{B}}T)=-\log\left[\rho(z)\right]$.
To check the efficacy of the technique we measured $V(z)$ for PLL-PEG coated colloids sedimented on PLL-PEG coated glass surface. As expected, the only potential attracting the bead towards the surface is the gravitational one, as shown in Fig.~\ref{figure_zpot}(a) curve {\it A}. Fitting the linear region of $V(z)$ we estimate the weight of the colloid $mg\approx2.6$~fN, which is in good agreement with the nominal value $\ge 3$~fN.
For plain silica colloids settled on the surface of a soft PAA gel, $V(z)$ reveals an additional attraction, as shown in Fig.~\ref{figure_zpot}(a) curve {\it B}. We quantify the attraction with a parabolic fit that gives an effective spring constant of $2.1\times10^{-9}$ N m$^{-1}$.  For the case of an ultra-soft gel, Fig.~\ref{figure_zpot}(a) curve {\it C} shows that the colloid-surface attraction is even stronger with an effective spring constant of $3.6\times10^{-9}$ N m$^{-1}$.\\
The attraction could lead to the partial inclusion of the beads into the gel surface \footnote{Diffusivity measurements of supported colloids demonstrate that inclusion cannot be complete.} thereby accentuating  a depletion-like interaction~\cite{Chen2005}, as sketched in Fig.~\ref{figure_pair}(b). Alternatively, a direct polymer bridging mediated by colloids-polymers surface adhesion can be responsible for the in-plane attraction, as shown in Fig.~\ref{figure_pair}(c)~\cite{Lafuma1991}.  Moreover, for the ultra-soft gels, the stronger colloid-gel attraction can result in a deeper penetration of the beads, which explains the stronger in-plane attraction either by increasing the excluded volume between two adjacent colloids or increasing the bead-gel contact surface. PLL-PEG coating on the colloids can reduce the polymer-colloid adhesion energy, thus the in-plane attraction, as found in $V(r)$ measurements. The correlation between in-plane and colloid-gel forces is sketched in Fig.~\ref{figure_zpot}(b) and (c). The repulsion found for $r\,\approx\,1.2\,\sigma$ can be explained as the elastic response of polymer coils squeezed between two adjacent beads. 
We also measured $V(z)$ on the soft and the ultra-soft gels using PLL-PEG coated beads, as shown in Fig.~\ref{figure_zpot} curves {\it B1} and {\it C1}. In the first case, steric stabilization does not affect $V(z)$ while in the case of ultra-soft gels the potential is slightly less attractive.\\ 
In summary, we demonstrated that silica colloids sedimented onto soft cross-linked polymer gel surfaces interact through a strong short-range non-hydrodynamic attractive potential that can be tuned either by changing the substrate stiffness or bead-surface properties, and, for the case of ultra-soft substrates, produces large scale aggregation, eventually leading to formation of ordered phases. The in-plane interaction correlates with a colloid-gel attraction that could result in a partial penetration of the colloids into the upper layers of the gels. All the experimental observations are consistent with the in-plane interactions arising from either depletion effects or direct polymer bridging. Further investigations are needed to determine which of these proposed mechanisms account for the aggregation of colloids on surfaces of soft polymer gels.\\

We thank Daan Frenkel for useful discussions. This work was supported by the Marie Curie Training Network ITN-COMPLOIDS No. 234810, the Ernest Oppenheimer and Schiff Fund, and the NSF Brandeis MRSEC.

\nocite{*}

\bibliography{soft_gels}

\begin{thebibliography}{27}%
\makeatletter
\providecommand \@ifxundefined [1]{%
 \@ifx{#1\undefined}
}%
\providecommand \@ifnum [1]{%
 \ifnum #1\expandafter \@firstoftwo
 \else \expandafter \@secondoftwo
 \fi
}%
\providecommand \@ifx [1]{%
 \ifx #1\expandafter \@firstoftwo
 \else \expandafter \@secondoftwo
 \fi
}%
\providecommand \natexlab [1]{#1}%
\providecommand \enquote  [1]{``#1''}%
\providecommand \bibnamefont  [1]{#1}%
\providecommand \bibfnamefont [1]{#1}%
\providecommand \citenamefont [1]{#1}%
\providecommand \href@noop [0]{\@secondoftwo}%
\providecommand \href [0]{\begingroup \@sanitize@url \@href}%
\providecommand \@href[1]{\@@startlink{#1}\@@href}%
\providecommand \@@href[1]{\endgroup#1\@@endlink}%
\providecommand \@sanitize@url [0]{\catcode `\\12\catcode `\$12\catcode
  `\&12\catcode `\#12\catcode `\^12\catcode `\_12\catcode `\%12\relax}%
\providecommand \@@startlink[1]{}%
\providecommand \@@endlink[0]{}%
\providecommand \url  [0]{\begingroup\@sanitize@url \@url }%
\providecommand \@url [1]{\endgroup\@href {#1}{\urlprefix }}%
\providecommand \urlprefix  [0]{URL }%
\providecommand \Eprint [0]{\href }%
\providecommand \doibase [0]{http://dx.doi.org/}%
\providecommand \selectlanguage [0]{\@gobble}%
\providecommand \bibinfo  [0]{\@secondoftwo}%
\providecommand \bibfield  [0]{\@secondoftwo}%
\providecommand \translation [1]{[#1]}%
\providecommand \BibitemOpen [0]{}%
\providecommand \bibitemStop [0]{}%
\providecommand \bibitemNoStop [0]{.\EOS\space}%
\providecommand \EOS [0]{\spacefactor3000\relax}%
\providecommand \BibitemShut  [1]{\csname bibitem#1\endcsname}%
\let\auto@bib@innerbib\@empty
\bibitem [{\citenamefont {Kepler}\ and\ \citenamefont
  {Fraden}(1994{\natexlab{a}})}]{Seth2}%
  \BibitemOpen
  \bibfield  {author} {\bibinfo {author} {\bibfnamefont {G.~M.}\ \bibnamefont
  {Kepler}}\ and\ \bibinfo {author} {\bibfnamefont {S.}~\bibnamefont
  {Fraden}},\ }\href {http://link.aps.org/doi/10.1103/PhysRevLett.73.356}
  {\bibfield  {journal} {\bibinfo  {journal} {Phys. Rev. Lett.}\ }\textbf
  {\bibinfo {volume} {73}},\ \bibinfo {pages} {356} (\bibinfo {year}
  {1994}{\natexlab{a}})}\BibitemShut {NoStop}%
\bibitem [{\citenamefont {Polin}\ \emph {et~al.}(2007)\citenamefont {Polin},
  \citenamefont {Grier},\ and\ \citenamefont {Han}}]{Grier2007}%
  \BibitemOpen
  \bibfield  {author} {\bibinfo {author} {\bibfnamefont {M.}~\bibnamefont
  {Polin}}, \bibinfo {author} {\bibfnamefont {D.~G.}\ \bibnamefont {Grier}}, \
  and\ \bibinfo {author} {\bibfnamefont {Y.}~\bibnamefont {Han}},\ }\href
  {http://link.aps.org/doi/10.1103/PhysRevE.76.041406} {\bibfield  {journal}
  {\bibinfo  {journal} {Phys. Rev. E}\ }\textbf {\bibinfo {volume} {76}},\
  \bibinfo {pages} {041406} (\bibinfo {year} {2007})}\BibitemShut {NoStop}%
\bibitem [{\citenamefont {Squires}\ and\ \citenamefont
  {Brenner}(2000)}]{Squires2000}%
  \BibitemOpen
  \bibfield  {author} {\bibinfo {author} {\bibfnamefont {T.~M.}\ \bibnamefont
  {Squires}}\ and\ \bibinfo {author} {\bibfnamefont {M.~P.}\ \bibnamefont
  {Brenner}},\ }\href {http://link.aps.org/doi/10.1103/PhysRevLett.85.4976}
  {\bibfield  {journal} {\bibinfo  {journal} {Phys. Rev. Lett.}\ }\textbf
  {\bibinfo {volume} {85}},\ \bibinfo {pages} {4976} (\bibinfo {year}
  {2000})}\BibitemShut {NoStop}%
\bibitem [{\citenamefont {Dom{\'\i}nguez}\ \emph {et~al.}(2005)\citenamefont
  {Dom{\'\i}nguez}, \citenamefont {Ottel},\ and\ \citenamefont
  {Dietrich}}]{Dietrich2005}%
  \BibitemOpen
  \bibfield  {author} {\bibinfo {author} {\bibfnamefont {A.}~\bibnamefont
  {Dom{\'\i}nguez}}, \bibinfo {author} {\bibfnamefont {M.}~\bibnamefont
  {Ottel}}, \ and\ \bibinfo {author} {\bibfnamefont {S.}~\bibnamefont
  {Dietrich}},\ }\href {http://stacks.iop.org/0953-8984/17/i=45/a=026}
  {\bibfield  {journal} {\bibinfo  {journal} {J. Phys.: Condens. Mat.}\
  }\textbf {\bibinfo {volume} {17}},\ \bibinfo {pages} {S3387} (\bibinfo {year}
  {2005})}\BibitemShut {NoStop}%
\bibitem [{\citenamefont {Im}\ \emph {et~al.}(2002)\citenamefont {Im},
  \citenamefont {Lim}, \citenamefont {Suh},\ and\ \citenamefont
  {Park}}]{Im2002}%
  \BibitemOpen
  \bibfield  {author} {\bibinfo {author} {\bibfnamefont {S.~H.}\ \bibnamefont
  {Im}}, \bibinfo {author} {\bibfnamefont {Y.~T.}\ \bibnamefont {Lim}},
  \bibinfo {author} {\bibfnamefont {D.~J.}\ \bibnamefont {Suh}}, \ and\
  \bibinfo {author} {\bibfnamefont {O.~O.}\ \bibnamefont {Park}},\ }\href
  {\doibase 10.1002/1521-4095(20021002)14:19<1367::AID-ADMA1367>3.0.CO;2-U}
  {\bibfield  {journal} {\bibinfo  {journal} {Adv. Mater.}\ }\textbf {\bibinfo
  {volume} {14}},\ \bibinfo {pages} {1367} (\bibinfo {year}
  {2002})}\BibitemShut {NoStop}%
\bibitem [{\citenamefont {Xia}\ \emph {et~al.}(2000)\citenamefont {Xia},
  \citenamefont {Gates}, \citenamefont {Yin},\ and\ \citenamefont
  {Lu}}]{Xia2000}%
  \BibitemOpen
  \bibfield  {author} {\bibinfo {author} {\bibfnamefont {Y.}~\bibnamefont
  {Xia}}, \bibinfo {author} {\bibfnamefont {B.}~\bibnamefont {Gates}}, \bibinfo
  {author} {\bibfnamefont {Y.}~\bibnamefont {Yin}}, \ and\ \bibinfo {author}
  {\bibfnamefont {Y.}~\bibnamefont {Lu}},\ }\href {\doibase
  10.1002/(SICI)1521-4095(200005)12:10<693::AID-ADMA693>3.0.CO;2-J} {\bibfield
  {journal} {\bibinfo  {journal} {Adv. Mater.}\ }\textbf {\bibinfo {volume}
  {12}},\ \bibinfo {pages} {693} (\bibinfo {year} {2000})}\BibitemShut
  {NoStop}%
\bibitem [{\citenamefont {Tarhan}\ and\ \citenamefont
  {Watson}(1996)}]{Tarhan1996}%
  \BibitemOpen
  \bibfield  {author} {\bibinfo {author} {\bibfnamefont {I.~I.~c.}\
  \bibnamefont {Tarhan}}\ and\ \bibinfo {author} {\bibfnamefont {G.~H.}\
  \bibnamefont {Watson}},\ }\href
  {http://link.aps.org/doi/10.1103/PhysRevLett.76.315} {\bibfield  {journal}
  {\bibinfo  {journal} {Phys. Rev. Lett.}\ }\textbf {\bibinfo {volume} {76}},\
  \bibinfo {pages} {315} (\bibinfo {year} {1996})}\BibitemShut {NoStop}%
\bibitem [{\citenamefont {Moshayedi}\ \emph {et~al.}(2010)\citenamefont
  {Moshayedi}, \citenamefont {da~F~Costa}, \citenamefont {Christ},
  \citenamefont {Lacour}, \citenamefont {Fawcett}, \citenamefont {Guck},\ and\
  \citenamefont {Franze}}]{Guck2010}%
  \BibitemOpen
  \bibfield  {author} {\bibinfo {author} {\bibfnamefont {P.}~\bibnamefont
  {Moshayedi}}, \bibinfo {author} {\bibfnamefont {L.}~\bibnamefont
  {da~F~Costa}}, \bibinfo {author} {\bibfnamefont {A.}~\bibnamefont {Christ}},
  \bibinfo {author} {\bibfnamefont {S.~P.}\ \bibnamefont {Lacour}}, \bibinfo
  {author} {\bibfnamefont {J.}~\bibnamefont {Fawcett}}, \bibinfo {author}
  {\bibfnamefont {J.}~\bibnamefont {Guck}}, \ and\ \bibinfo {author}
  {\bibfnamefont {K.}~\bibnamefont {Franze}},\ }\href
  {http://stacks.iop.org/0953-8984/22/i=19/a=194114} {\bibfield  {journal}
  {\bibinfo  {journal} {J. Phys.: Condens. Mat.}\ }\textbf {\bibinfo {volume}
  {22}},\ \bibinfo {pages} {194114} (\bibinfo {year} {2010})}\BibitemShut
  {NoStop}%
\bibitem [{\citenamefont {Discher}\ \emph {et~al.}(2005)\citenamefont
  {Discher}, \citenamefont {Janmey},\ and\ \citenamefont {Wang}}]{Discher2005}%
  \BibitemOpen
  \bibfield  {author} {\bibinfo {author} {\bibfnamefont {D.~E.}\ \bibnamefont
  {Discher}}, \bibinfo {author} {\bibfnamefont {P.}~\bibnamefont {Janmey}}, \
  and\ \bibinfo {author} {\bibfnamefont {Y.-l.}\ \bibnamefont {Wang}},\ }\href
  {http://www.sciencemag.org/content/310/5751/1139.abstract} {\bibfield
  {journal} {\bibinfo  {journal} {Science}\ }\textbf {\bibinfo {volume}
  {310}},\ \bibinfo {pages} {1139} (\bibinfo {year} {2005})}\BibitemShut
  {NoStop}%
\bibitem [{\citenamefont {Engler}\ \emph {et~al.}(2004)\citenamefont {Engler},
  \citenamefont {Bacakova}, \citenamefont {Newman}, \citenamefont {Hategan},
  \citenamefont {Griffin},\ and\ \citenamefont {Discher}}]{Engler2004}%
  \BibitemOpen
  \bibfield  {author} {\bibinfo {author} {\bibfnamefont {A.}~\bibnamefont
  {Engler}}, \bibinfo {author} {\bibfnamefont {L.}~\bibnamefont {Bacakova}},
  \bibinfo {author} {\bibfnamefont {C.}~\bibnamefont {Newman}}, \bibinfo
  {author} {\bibfnamefont {A.}~\bibnamefont {Hategan}}, \bibinfo {author}
  {\bibfnamefont {M.}~\bibnamefont {Griffin}}, \ and\ \bibinfo {author}
  {\bibfnamefont {D.}~\bibnamefont {Discher}},\ }\href
  {http://www.sciencedirect.com/science/article/B94RW-4V352BY-28/2/86006418cc4324f90355606a2036a74a}
  {\bibfield  {journal} {\bibinfo  {journal} {Biophys. J.}\ }\textbf {\bibinfo
  {volume} {86}},\ \bibinfo {pages} {617} (\bibinfo {year} {2004})}\BibitemShut
  {NoStop}%
\bibitem [{\citenamefont {Ulrich}\ \emph {et~al.}(2009)\citenamefont {Ulrich},
  \citenamefont {de~Juan~Pardo},\ and\ \citenamefont {Kumar}}]{Ulrich2009}%
  \BibitemOpen
  \bibfield  {author} {\bibinfo {author} {\bibfnamefont {T.~A.}\ \bibnamefont
  {Ulrich}}, \bibinfo {author} {\bibfnamefont {E.~M.}\ \bibnamefont
  {de~Juan~Pardo}}, \ and\ \bibinfo {author} {\bibfnamefont {S.}~\bibnamefont
  {Kumar}},\ }\href
  {http://cancerres.aacrjournals.org/content/69/10/4167.abstract} {\bibfield
  {journal} {\bibinfo  {journal} {Cancer Res.}\ }\textbf {\bibinfo {volume}
  {69}},\ \bibinfo {pages} {4167} (\bibinfo {year} {2009})}\BibitemShut
  {NoStop}%
\bibitem [{\citenamefont {Zajac}\ and\ \citenamefont
  {Discher}(2008)}]{Zajac2008}%
  \BibitemOpen
  \bibfield  {author} {\bibinfo {author} {\bibfnamefont {A.~L.}\ \bibnamefont
  {Zajac}}\ and\ \bibinfo {author} {\bibfnamefont {D.~E.}\ \bibnamefont
  {Discher}},\ }\href
  {http://www.sciencedirect.com/science/article/B6VRW-4TS80VG-2/2/769c728e73998b6e37127d070dc75f11}
  {\bibfield  {journal} {\bibinfo  {journal} {Curr. Opin. Cell Biol.}\ }\textbf
  {\bibinfo {volume} {20}},\ \bibinfo {pages} {609} (\bibinfo {year}
  {2008})}\BibitemShut {NoStop}%
\bibitem [{Note1()}]{Note1}%
  \BibitemOpen
  \bibinfo {note} {Measured using Physica MCR bulk rheometer, Anton Paar,
  Austria.}\BibitemShut {Stop}%
\bibitem [{\citenamefont {Crocker}\ and\ \citenamefont
  {Grier}(1994)}]{Crocker1994}%
  \BibitemOpen
  \bibfield  {author} {\bibinfo {author} {\bibfnamefont {J.~C.}\ \bibnamefont
  {Crocker}}\ and\ \bibinfo {author} {\bibfnamefont {D.~G.}\ \bibnamefont
  {Grier}},\ }\href {http://link.aps.org/doi/10.1103/PhysRevLett.73.352}
  {\bibfield  {journal} {\bibinfo  {journal} {Phys. Rev. Lett.}\ }\textbf
  {\bibinfo {volume} {73}},\ \bibinfo {pages} {352} (\bibinfo {year}
  {1994})}\BibitemShut {NoStop}%
\bibitem [{\citenamefont {Crocker}\ and\ \citenamefont
  {Grier}(1996{\natexlab{a}})}]{Crocker1996b}%
  \BibitemOpen
  \bibfield  {author} {\bibinfo {author} {\bibfnamefont {J.~C.}\ \bibnamefont
  {Crocker}}\ and\ \bibinfo {author} {\bibfnamefont {D.~G.}\ \bibnamefont
  {Grier}},\ }\href
  {http://www.sciencedirect.com/science/article/pii/S0021979796902179}
  {\bibfield  {journal} {\bibinfo  {journal} {J. Colloid Interf. Sci.}\
  }\textbf {\bibinfo {volume} {179}},\ \bibinfo {pages} {298} (\bibinfo {year}
  {1996}{\natexlab{a}})}\BibitemShut {NoStop}%
\bibitem [{Note2()}]{Note2}%
  \BibitemOpen
  \bibinfo {note} {Home-made optical tweezers based on a 2W 1064nm laser
  (CrystaLaser) mounted on a Nikon Eclipse Ti-E microscope with a Plan Apo VC
  60x WI 1.2 N.A. objective and a Pike F-100B CCD camera (Allied Vision
  Technologies).}\BibitemShut {Stop}%
\bibitem [{Note3()}]{Note3}%
  \BibitemOpen
  \bibinfo {note} {We developed our customized software based on open source
  MATLAB routines available at the URL \protect \url
  {http://physics.georgetown.edu/matlab}}\BibitemShut {NoStop}%
\bibitem [{\citenamefont {Baumgartl}\ and\ \citenamefont
  {Bechinger}(2005)}]{Baumgartl2005}%
  \BibitemOpen
  \bibfield  {author} {\bibinfo {author} {\bibfnamefont {J.}~\bibnamefont
  {Baumgartl}}\ and\ \bibinfo {author} {\bibfnamefont {C.}~\bibnamefont
  {Bechinger}},\ }\href {http://stacks.iop.org/0295-5075/71/i=3/a=487}
  {\bibfield  {journal} {\bibinfo  {journal} {Europhys. Lett.}\ }\textbf
  {\bibinfo {volume} {71}} (\bibinfo {year} {2005})}\BibitemShut {NoStop}%
\bibitem [{Note4()}]{Note4}%
  \BibitemOpen
  \bibinfo {note} {Tracking the distance between plain colloids on ultra-soft
  gels was not reliable due to rapid aggregation}\BibitemShut {NoStop}%
\bibitem [{Note5()}]{Note5}%
  \BibitemOpen
  \bibinfo {note} {For practical reasons our samples could not be sealed. The
  resulting slow evaporation could possibly increase the buffer concentration
  leading to further reduction in the Debye screening length. No measurable
  flow at the level of the surface was observed.}\BibitemShut {Stop}%
\bibitem [{\citenamefont {Crocker}\ and\ \citenamefont
  {Grier}(1996{\natexlab{b}})}]{Crocker1996}%
  \BibitemOpen
  \bibfield  {author} {\bibinfo {author} {\bibfnamefont {J.~C.}\ \bibnamefont
  {Crocker}}\ and\ \bibinfo {author} {\bibfnamefont {D.~G.}\ \bibnamefont
  {Grier}},\ }\href {http://link.aps.org/doi/10.1103/PhysRevLett.77.1897}
  {\bibfield  {journal} {\bibinfo  {journal} {Phys. Rev. Lett.}\ }\textbf
  {\bibinfo {volume} {77}},\ \bibinfo {pages} {1897} (\bibinfo {year}
  {1996}{\natexlab{b}})}\BibitemShut {NoStop}%
\bibitem [{\citenamefont {Asakura}\ and\ \citenamefont
  {Oosawa}(1954)}]{ASAKURA1954}%
  \BibitemOpen
  \bibfield  {author} {\bibinfo {author} {\bibfnamefont {S.}~\bibnamefont
  {Asakura}}\ and\ \bibinfo {author} {\bibfnamefont {F.}~\bibnamefont
  {Oosawa}},\ }\href@noop {} {\bibfield  {journal} {\bibinfo  {journal} {J.
  Chem. Phys.}\ }\textbf {\bibinfo {volume} {22}},\ \bibinfo {pages} {1255}
  (\bibinfo {year} {1954})}\BibitemShut {NoStop}%
\bibitem [{\citenamefont {Asakura}\ and\ \citenamefont
  {Oosawa}(1958)}]{ASAKURA1958}%
  \BibitemOpen
  \bibfield  {author} {\bibinfo {author} {\bibfnamefont {S.}~\bibnamefont
  {Asakura}}\ and\ \bibinfo {author} {\bibfnamefont {F.}~\bibnamefont
  {Oosawa}},\ }\href@noop {} {\bibfield  {journal} {\bibinfo  {journal} {J.
  Polym. Sci.}\ }\textbf {\bibinfo {volume} {33}},\ \bibinfo {pages} {183}
  (\bibinfo {year} {1958})}\BibitemShut {NoStop}%
\bibitem [{\citenamefont {Chen}\ and\ \citenamefont {Ma}(2005)}]{Chen2005}%
  \BibitemOpen
  \bibfield  {author} {\bibinfo {author} {\bibfnamefont {K.}~\bibnamefont
  {Chen}}\ and\ \bibinfo {author} {\bibfnamefont {Y.-q.}\ \bibnamefont {Ma}},\
  }\href {\doibase 10.1021/jp051403u} {\bibfield  {journal} {\bibinfo
  {journal} {J. Phys. Chem. B}\ }\textbf {\bibinfo {volume} {109}},\ \bibinfo
  {pages} {17617} (\bibinfo {year} {2005})}\BibitemShut {NoStop}%
\bibitem [{\citenamefont {Kepler}\ and\ \citenamefont
  {Fraden}(1994{\natexlab{b}})}]{Seth1}%
  \BibitemOpen
  \bibfield  {author} {\bibinfo {author} {\bibfnamefont {G.~M.}\ \bibnamefont
  {Kepler}}\ and\ \bibinfo {author} {\bibfnamefont {S.}~\bibnamefont
  {Fraden}},\ }\href {\doibase 10.1021/la00020a003} {\bibfield  {journal}
  {\bibinfo  {journal} {Langmuir}\ }\textbf {\bibinfo {volume} {10}},\ \bibinfo
  {pages} {2501} (\bibinfo {year} {1994}{\natexlab{b}})}\BibitemShut {NoStop}%
\bibitem [{Note6()}]{Note6}%
  \BibitemOpen
  \bibinfo {note} {Diffusivity measurements of supported colloids demonstrate
  that inclusion cannot be complete.}\BibitemShut {Stop}%
\bibitem [{\citenamefont {Lafuma}\ \emph {et~al.}(1991)\citenamefont {Lafuma},
  \citenamefont {Wong},\ and\ \citenamefont {Cabane}}]{Lafuma1991}%
  \BibitemOpen
  \bibfield  {author} {\bibinfo {author} {\bibfnamefont {F.}~\bibnamefont
  {Lafuma}}, \bibinfo {author} {\bibfnamefont {K.}~\bibnamefont {Wong}}, \ and\
  \bibinfo {author} {\bibfnamefont {B.}~\bibnamefont {Cabane}},\ }\href
  {http://www.sciencedirect.com/science/article/pii/0021979791904339}
  {\bibfield  {journal} {\bibinfo  {journal} {J. Colloid Interf. Sci.}\
  }\textbf {\bibinfo {volume} {143}},\ \bibinfo {pages} {9} (\bibinfo {year}
  {1991})}\BibitemShut {NoStop}%
\end{thebibliography}%

\end{document}